\documentclass[english,aps,twocolumn,notitlepage,tightenlines,nofootinbib]{revtex4}
\usepackage[T1]{fontenc}
\usepackage[latin9]{inputenc}
\usepackage{verbatim}
\usepackage{amsmath}
\usepackage{dcolumn}
\usepackage{graphicx}
\newcommand{\be}{\begin{equation}}
\newcommand{\ee}{\end{equation}}
\newcommand{\bea}{\begin{eqnarray}}
\newcommand{\eea}{\end{eqnarray}}
\newcommand{\ep}{\epsilon}

\usepackage{babel}
\usepackage{color}
\begin{document}
\global\long\def\order#1{\mathcal{O}\left(#1\right)}
\global\long\def\d{\mathrm{d}}
\global\long\def\P{P}
\global\long\def\amp{{\mathcal M}}

\title{ 
\vskip-1cm{\baselineskip14pt
    \begin{flushright}
      \normalsize TTP15-009
  \end{flushright}}
Production of two Z-bosons in  gluon fusion 
in the heavy top quark approximation
}

\author{Kirill Melnikov}
\email{melnikov@pha.jhu.edu}
\affiliation{ Institute for Theoretical Particle Physics, Karlsruhe Institute of Technology, 
Karlsruhe, Germany}

\author{Matthew Dowling }
\email{matthew.dowling@kit.edu}
\affiliation{ Institute for Theoretical Particle Physics, Karlsruhe Institute of Technology, 
Karlsruhe, Germany}

\begin{abstract}
We compute  QCD radiative corrections to the continuum production of 
a pair of $Z$-bosons in the annihilation of two gluons.  We only consider 
the contribution of the top quark loops  and we treat them  assuming that 
$m_t$ is much larger than any other kinematic invariant in the problem. 
We estimate the QCD corrections to $pp \to ZZ$  using the  first non-trivial term in the expansion 
in the inverse top quark mass and we compare them  to  QCD corrections of the 
signal process, $pp \to H \to ZZ$. 
\end{abstract}

\maketitle

\section{Introduction} 
Production of pairs of vector bosons in proton collisions is   one of the most interesting processes  
 studied at the LHC at the end of the Run I \cite{atlas7,cms7,cms8}.
  Indeed, $pp \to ZZ$, $pp \to  W^+W^-$, 
and $pp \to \gamma \gamma$,  play an important role  in Higgs boson physics, provide stringent tests 
of the Standard Model and give constraints on anomalous electroweak triple gauge boson couplings. 
In the case of Higgs physics, such processes are essential for understanding backgrounds to Higgs 
boson signals, for    constraining   anomalous Higgs boson couplings, for measuring the 
quantum numbers of the Higgs boson and for studying the Higgs boson width, see e.g. 
Refs.\cite{Khachatryan:2014jba,Khachatryan:2014kca,Khachatryan:2014iha,atlaswidth}.

Production of electroweak gauge boson pairs occurs mainly due to quark-antiquark annihilation. 
This contribution is known through next-to-next-to-leading order (NNLO) in perturbative QCD
\cite{Cascioli:2014yka,tg1}. However, as was pointed out in 
Refs.\cite{Glover:1988rg,Glover:1988fe,Dicus:1987dj}, 
there is a sizeable contribution of the gluon annihilation channel 
$gg \to V_1 V_2$ whose significance depends on the selection  cuts.  For example ~\cite{Binoth:2006mf}, 
agressive cuts applied to $pp \to W^+W^-$ to separate the Higgs boson signal from the continuum 
background can increase 
the fraction of gluon fusion events in the background sample to ${\cal O}(30)$ percent.
Since $gg \to V_1 V_2$ is the one-loop process and since production of 
electroweak boson pairs at leading order occurs only in the $q \bar q$ channel, the gluon fusion contribution 
to $pp \to V_1 V_2$ through NNLO  only needs to be known at the leading, one-loop, approximation. 
Thus, all existing numerical  estimates of the significance of the
gluon fusion mechanism in weak boson pair production ignore radiative corrections 
to $gg \to ZZ$ that are, potentially,  quite large~\cite{Bonvini:2013jha}.  The need 
to have a more accurate estimate of QCD corrections to gluon fusion processes was strongly 
emphasized in Ref.~\cite{atlaswidth}, in the context of the Higgs width and generic off-shell 
measurements \cite{Caola:2013yja,
Campbell:2013una, ellis,Azatov:2014jga}.

The largest contribution to $gg \to V_1 V_2$ comes from quarks of the first two generations 
that can be taken to be massless (for a  recent discussion, see  Ref.~\cite{ellis}).
The contribution of the third generation is, in general 
smaller. For example, in the case of $W^+W^-$ production it is known that the third generation 
changes the $gg \to V_1 V_2$ production cross-section by ${\cal O}(10)$ percent at the 
$13~{\rm TeV}$ LHC \cite{ellis}.   Since gluon fusion contributes ${\cal O}(5)$ percent   
to the $pp \to W^+W^-$ cross-section,  the impact of top quark loops on the cross-section is marginal. 
On the other hand, studies of {\it off-shell} 
Higgs boson production  may be senstitive to the third 
generation  of quarks and, especially, to {\it  massive top quark loops}. Of particular concern in this context 
is the  interference of $gg \to V_1 V_2$  and $gg \to H^* \to V_1 V_2$  amplitudes, 
as discussed recently in Refs.~\cite{ellis,Azatov:2014jga}.

The recent progress in calculating  two-loop integrals with two massless and two massive external lines 
\cite{Gehrmann:2014bfa,nonplanar,planar,Papadopoulos:2014hla} 
enables computation of  scattering amplitudes and, eventually, QCD corrections 
to the production of pairs of vector bosons in gluon fusion through loops of {\it massless} quarks. 
A similar progress towards computing $gg \to V_1 V_2$ contributions mediated by loops of massive 
quarks is very desirable but, probably, it will not be immediate 
since two-loop computations of four-point functions with internal massive lines are beyond 
 existing technical capabilities. 

In this situation, it is  useful  to think about alternative approaches that will allow 
an  estimate   of QCD radiative  corrections to gluon fusion processes 
mediated by heavy quark loops.
A practical opportunity   is provided by the expansion of amplitudes in the inverse quark mass. 
Indeed, this approach reduces the calculation of the one-loop $gg \to ZZ$ amplitude 
with massive internal particles to the calculation of tadpole diagrams which makes generalization 
to higher-order corrections relatively straightforward.\footnote{A similar approach is more problematic 
in the case of $gg \to W^+W^-$ where the third generation loops contain both massive (top)  and massless (bottom) quarks.}
 While the expansion of cross-sections in $1/m_t$ 
cannot be fully justified, particularly for  large invariant masses of $Z$-pairs, we have significant 
evidence that such computations do provide a reasonable estimate of the size of QCD corrections. 
Indeed, this is an approach that is taken in calculations of single- \cite{shiggs} and double-Higgs 
\cite{dhiggs} production 
at the LHC where {\it exact} one-loop computations supplemented with QCD corrections 
calculated  in the $m_t \to \infty$ approximation are believed  to provide reasonably accurate descriptions 
of  these processes for realistic values 
of  top quark and Higgs boson masses. 
It is clear that a similar approach should be applicable to the production of pairs of 
$Z$- bosons in gluon fusion through the top quark loop. In fact, the $m_t \to \infty$ 
approximation for  $gg \to ZZ$ should work better than for the case of  Higgs pair production 
since $2m_Z$ is  smaller than $2 m_H$.  

The goal of this paper is to make  the first step 
towards estimating the NLO QCD correction to the production of $Z$-boson pairs  in gluon 
fusion.  To this end, we take {\it continuum}  production of two on-shell $Z$-bosons in gluon fusion 
through the top quark loop and compute the NLO QCD corrections to it in the heavy top approximation. 
This allows us to compare, for the first time, the QCD corrections to the ``background'' $gg \to ZZ$ and 
the signal $gg \to H \to ZZ$ processes.  We find that the corrections to the two processes are 
indeed similar, in accord with the arguments in Ref.~\cite{Bonvini:2013jha}. 

It should be clear from the previous discussion that computation of QCD corrections to 
the total cross-section is just one of many 
interesting physics questions, 
including 
interference with the Higgs signal on and off the mass shell,  
combination of light  and heavy quark contirbutions to the 
$gg \to ZZ$ amplitude, estimates of $1/m_t$ corrections to 
cross-sections etc., that can  be discussed in the context of vector  
boson pair production in gluon fusion,
once the two-loop amplitude for $gg \to ZZ$ becomes available.  
We plan to address  these questions  in the near future. 

The paper is organized as follows. In Section~\ref{section1}, we describe  the general 
set up of the computation and present the analytic result for the two-loop amplitude $gg \to ZZ$ 
in the large-$m_t$ approximation. 
In Section~\ref{section2}, we derive the  analytic formulas for 
$gg \to ZZ$ partonic cross-sections. In Section~\ref{section3} we discuss  numerical results.
We present our conclusions in Section~\ref{section4}.

\section{The set up of the computation} 
\label{section1}

We consider the process $g(p_1) + g(p_2) \to Z(p_3) + Z(p_4)$ in a theory where $Z$-bosons 
only couple to top quarks.\footnote{Such a theory is anomalous and, in principle,  one should carefully 
consider diagrams where a $Z$-boson couples to {\it two} gluons.  Given the order in the $1/m_t$ 
expansion that we work to in this paper, Feynman diagrams where each $Z$ independently couples to 
gluon pairs  do not contribute.}
Contributions of massless quarks are not included in the computation 
except  in the running of the coupling constant where the complete $\beta$-function is employed. 
The coupling of top quarks to $Z$-bosons is given by a linear combination 
of vector and axial couplings
\be
Z \bar t  t \in -i \gamma^\mu ( g_V + g_A \gamma_5),
\ee
where $g_V = e/(2 \sin 2 \theta_W) ( 1- 8/3 \sin^2 \theta_W) $ and 
$g_A  = e/(2 \sin 2 \theta_W)$.

\begin{figure}[t]
  \centering
  \includegraphics[angle=0,width=0.4\textwidth]{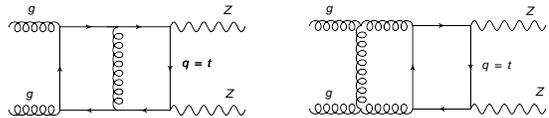}\\
  \caption{Representative two-loop diagrams that describe  production of Z-boson pairs 
in gluon fusion. }
  \label{fig0}
\end{figure}

The scattering amplitude  for $gg \to ZZ$ can be written as a sum of axial, vector and 
mixed terms
\be
{\cal A}_{gg \to ZZ}  = i a_s \delta^{a_1 a_2}
\left ( g_A^2 {\cal A}^{aa} + g_V^2 {\cal A}^{vv}  + g_A g_V {\cal A}^{av} \right ),
\label{eq1}
\ee
where $a_{1,2}$ are the color indices of the colliding gluons, 
$$
a_s =  \frac{\Gamma(1+\ep)}{(4\pi)^{-\ep}} \; \frac{ \alpha_s(\mu)}{\pi },
$$ 
and $\alpha_s(\mu)$ is the $\overline {\rm MS}$ QCD coupling constant in the theory with five active 
flavors.\footnote{ The contributions to the running of the coupling constant due to top quarks 
are subtracted at zero external momentum, i.e. on the  mass-shell of an  external gluon.}
We note that thanks to charge parity conservation, the axial-vector term vanishes, i.e. $A^{av} = 0$. 
The remaining two terms --  axial-axial ${\cal A}^{aa} $ 
and vector-vector ${\cal A}^{vv}$  --  do not vanish but they 
behave differently under the $1/m_t$ expansion.

Indeed, consider  a vector-current interaction of $Z$-bosons with top quarks.
The $gg \to ZZ$ amplitude behaves as ${\cal A}^{vv} \sim s^2/m_t^4$. This is a direct consequence 
of the vector current conservation which requires that, in the 
expression for the amplitude, each  polarization vector for 
either a gluon or an electroweak gauge boson is  accompanied by its   momentum. A similar suppression 
in the QED case is familiar in the context of Euler-Heisenberg Lagrangian.

However, if  the interaction of $Z$-bosons with top quarks is mediated 
by the axial current, the situation is different since the axial current is not conserved.  
As a consequence, the scattering amplitude can involve two polarization vectors of the 
$Z$-bosons {\it without} corresponding momenta while the gluon polarization vectors should still 
be accompanied by their momenta to satisfy the vector current conservation constraint.  Hence, 
we expect that  the axial amplitude ${\cal A}^{aa}$ behaves as ${\cal A}^{aa} \sim s/m_t^2$ and, therefore, 
exhibits weaker  suppression in the $m_t \to \infty$ 
limit   compared to ${\cal A}^{vv}$. 
Since our goal in this paper is to study the leading term of the ${\cal A}_{gg \to ZZ}$ amplitude 
in the $m_t \to \infty$ expansion, we conclude that we only need to study terms induced by the axial 
coupling of the $Z$-bosons to top quarks. 

The production of $Z$-boson pairs in gluon fusion is a loop-induced process. 
There are eight one-loop and ninty-three two-loop diagrams that contribute 
to  $gg \to ZZ$.  Some examples are shown in Fig.~\ref{fig0}.  We compute these diagrams 
using asymptotic expansions in the inverse top quark mass \cite{Smirnov:2013}. The essense 
of this procedure is that the loop momenta in each of the Feynman diagrams 
are separated into soft $l \sim p_{1,..4}$ and  hard $l \sim m_t$. All possible 
assignments must be  considered. The integrand of a Feynman diagram is then Taylor 
expanded in all quantities that are considered small. Upon such an expansion, computation of Feynman 
diagrams significantly simplifies.  Consider one-loop diagrams as an example. 
In this case the momentum can only be  hard, 
$l \sim m_t$, 
 and so integrands for 
all diagrams are expanded in Taylor series in their external momenta.\footnote{If the loop momentum 
is assumed to be soft, each propagator is  expanded in $l/m_t$ generating scaleless integrals.}
  All one-loop integrals  then become vacuum tadpole integrals and it is straightforward to 
evaluate them.  

The situation with two-loop integrals is similar although somewhat more involved. 
Indeed, in this case   two momentum configurations 
are possible: either both loop momenta are  
hard or one of the loop momenta is  hard and the other one is soft.  If both loop momenta 
are hard, the calculation is reduced to the calculation of two-loop vacuum tadpole 
diagrams.  If one of the loop momenta is soft and the other one is hard, the diagram 
factorizes  into a product of   one-loop integrals, the most complicated 
of which is a three-point function  with all internal and  two external lines massless.  

We will now present our results for the  $gg \to ZZ$ amplitude.
We write it 
as an expansion in the strong coupling constant 
\be
{\cal A}^{aa} = \frac{1}{3 m_t^2}  \left(\frac{\mu}{m_t}  \right )^{2\ep} 
\left \{ {\cal A}^{aa}_{1} + a_s \; \left(\frac{\mu}{m_t}  \right )^{2\ep}   {\cal A}^{aa}_{2} \right \}.
\ee
To emphasize constraints on the amplitude that follow from 
gauge invariance, we introduce  the Fourier transform of the 
field-strength tensor for each of the gluons
\be
f^{i,\mu \nu} = p_i^{\mu}  \epsilon_i^{\nu} -  p_i^{\nu} \epsilon_i^{\mu},\;\;\;\;i =1,2.
\ee
The one-loop amplitude reads 
\be
\begin{split} 
{\cal A}^{aa}_{1} =  (1+\ep) \left ( f_{\mu \rho}^{1} f^{2,\mu}_{\beta} 
- \frac{g_{\rho \beta}}{2} f_{\mu \rho}^{1} f^{2,\mu \rho}   \right ) t_{34}^{\rho \beta},
\label{eq7}
\end{split}
\ee
where 
\be
t_{34}^{\rho \beta} = \epsilon_3^{\rho} \epsilon_{4}^{\beta} 
 + \epsilon_{4}^{ \rho} \epsilon_{3}^{\beta},
\ee
and $\epsilon_{3,4}$ are the polarization vectors of the two $Z$-bosons. 
We emphasize that the dependence of the amplitude on the dimensional regularization  
parameter $\ep$ in Eq.(\ref{eq7}) is exact.

The two-loop amplitude reads
\be
\begin{split} 
 {\cal A}^{aa,2}  & =  \left (
 - \left ( \frac{3}{2\ep^2}  + \frac{\beta_0}{2 \ep} \right ) \left ( \frac{-s-i0}{m_t^2} \right )^{-\ep}
\right. 
\\
& \left.  -\frac{\beta_0}{2} L_{s\mu} + \frac{11}{4} L_{sm} + \frac{\pi^2}{4} - \frac{175}{36} 
\right ) {\cal A}^{aa,1}
\\
& 
+\frac{1}{2} f_{\mu \rho}^{1} f^{2,\mu \rho}  t_{34 \beta}^{\beta} \left (-\frac{385}{72}
  + \frac{11}{8} L_{sm}  \right ) 
\\
& - \frac{1}{2s} f_{\mu \nu}^{1} f^{2,\mu \nu}  t_{34}^{\rho, \beta} (p_{1,\rho}p_{1,\beta} + p_{2,\rho} p_{2, \beta} )
\\
&
+ \frac{3}{2s} f_{\mu \rho}^{1} p_1^\mu f^{2}_{\nu \beta} p_2^{\nu} t_{34}^{\rho, \beta} + \mathcal{O}(\epsilon),
\end{split} 
\ee
where $L_{s\mu} = \log((-s-i0)/\mu^2)$ and $L_{sm} = \log((-s-i0)/m_t^2)$ and $\beta_0 = 11/2 - N_f/3  $ with $N_f=5$ 
being the number of {\it massless} fermions.

In addition to virtual corrections,  we  require an amplitude for the real emission process, 
$g(p_1)g(p_2) \to Z(p_3) + Z(p_4) + g(p_5) $.
To order $1/m_t^2$ the corresponding amplitude can, in principle, be obtained from the amplitude of the one-loop scattering 
process in Eq.(\ref{eq7}), if the latter is written as a term in an effective Lagrangian, and then used to generate 
amplitudes with additional gluons in the final state.  
However, 
it is also convenient to apply the asymptotic expansion procedure to the computation of the relevant 
diagrams since this approach can be used to obtain the amplitude for $gg \to ZZ+g$ beyond the leading 
order in $1/m_t$.   

We use the second approach to compute the $gg \to ZZ+g$ amplitude. There are fifty diagrams that 
contribute to this process and we compute the relevant  diagrams using the $1/m_t$ expansion. The technical 
details of the calculation are identical to the calculation of the scattering amplitude for the 
$gg \to ZZ$ process and we do not repeat it here.   Unfortunately, the expresssion for the amplitude 
appears to be too complex to be presented here. 

To calculate the production cross-section, we square the elastic and the inelastic scattering 
amplitudes and integrate them over the corresponding phase-spaces. In order to make the cross-section 
finite, we need to remove collinear singularities by performing renormalization of parton distribution 
functions. All of these steps are relatively standard and well-known; for this reason we refrain from 
describing them in detail. 

\section{Production cross-section} 
\label{section2}

We are now in position to present  results for the gluon fusion contribution 
to the production cross-section $pp \to ZZ$. As explained previously, we only 
consider loops of top quarks and we work to leading order in the $1/m_t$ expansion. 
We take  the invaraint mass of the $Z$-boson pair to be  $q^2$ and write the 
differential cross-section as 
a  convolution of the partonic production 
cross-section and the parton distribution functions
\be
\begin{split}
\frac{{\rm d} \sigma_{pp \to ZZ} }{{\rm d} q^2}  & = 
\int \limits_{0}^{1} {\rm d} x_1 {\rm d} x_2 {\rm d} z \; f_g(x_1) f_g(x_2) 
\\
& 
\times \delta\left ( z - \frac{\tau}{x_1 x_2} \right )
  \frac{{\rm d} \sigma_{gg \to ZZ}}{{\rm d } q^2}(s,q^2)|_{s = q^2/z}. 
\end{split} 
\label{eq8}
\ee
In Eq.(\ref{eq8}), we used the following notation:  $f_{g}(x_{1,2})$ are the 
gluon parton distribution functions, $\tau = q^2/S_{\rm hadr}$ and 
$S_{\rm hadr}$ is the hadronic center-of-mass energy squared. 
We note that dependencies on the renormalization 
and factorization scales in Eq.(\ref{eq8}) are  suppressed.  
In what follows, we take the factorization and the renormalization scales 
to be equal. 

It is conventional to parametrize 
the partonic cross-section as 
\be
q^2\frac{{\rm d} \sigma_{gg \to ZZ}}{{\rm d } q^2}(s,q^2)|_{s = q^2/z} = \sigma_0 z G(z,q^2),
\ee
where 
\be
\sigma_0 = \frac{g_A^4  q^2 }{2^{10} \pi m_t^4}
\left ( \frac{\alpha_s(\mu)}{\pi} \right  )^2  \sqrt{1-\frac{4m_Z^2}{q^2}},   
\ee
and $G(z,q^2)$ can be written as series in the strong coupling constant. To present it, we 
introduce a parameter $r$ defined as  $r = q^2/(4m_Z^2)$.  We find 
\be
\begin{split}
& G(z,q^2)  = 
\Bigg [ \Delta_0 \delta(1-z) 
+ a_s  \Big (   
 \Delta_V \delta(1-z) + 
\\
& 
6\Delta_0   \left (  2D_1(z) +\ln \frac{q^2}{\mu^2} D_0(z) \right )
+ \Delta_H
\Big  ) \Bigg ],
\end{split}
\ee
where $D_i(z) = \left [ \ln(1-z)^i/(1-z) \right ]_+$ are the different plus-distribution functions  and 
\be
\label{eqdelta0}
\begin{split}
& \Delta_0 = \frac{73}{270} - \frac{2r}{15} + \frac{34 r^2}{135}.
\end{split}
\ee
We note that $\Delta_0$ has a strong dependence on $q^2$. The  leading growth caused by the ${\cal O}(r^2) \sim q^4/m_Z^4$ 
term in Eq.(\ref{eqdelta0}) 
is the consequence of the fact that pairs of longitudinal bosons can be produced. It is this growth 
that should, eventually, get tamed by the destructive interference of $gg \to ZZ$ and $gg \to H^* \to ZZ$ 
amplitudes. 

The virtual corrections combined with finite parts of soft emissions read 
\be
\begin{split}
& \Delta_V = \frac{2473 - 8661 r + 5798 r^2}{2430} 
\\
& + \frac{ ( 73 - 36 r + 68 r^2 ) \pi^2}{270} 
+ \frac{11(7+6r + 2r^2)}{135} \ln \frac{q^2}{m_t^2}.
\end{split}
\ee 
The contributions of hard emissions, not proportional to the leading order cross-section read
\be
\begin{split}
\Delta_H & = 
    \frac{6\Delta_0}{z} 
\left (  (\omega(z) - z\kappa(z)) 
\ln 
\left (  \frac{q^2 (1-z)^2 }{\mu^2 } \right )  
\right. 
\\
& \left.  - \omega(z)^2 \frac{\ln(z)}{(1-z)} \right )
    + (1-z) \Bigg [ \frac{r (11 \kappa(z) -46 z)}{15 z}
\\
&
     - \frac{r^2 (187 \kappa(z) -302z)}{135 z}- \frac{(803 \kappa(z)-598z)}{540 z}
\Bigg ],
\end{split}
\ee
where $\omega(z) = 1-z+z^2$ and $\kappa(z) = 1+z^2$.

\begin{figure}[t]
  \centering
  \includegraphics[angle=0,width=0.45\textwidth]{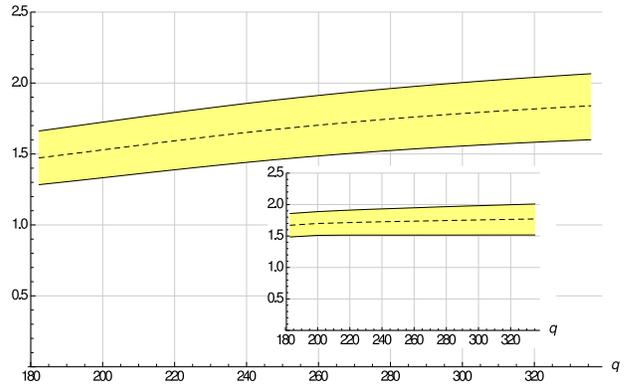}\;\;
  \caption{Main plot: NLO $K$-factor for $gg \to ZZ$ production through the top quark loop 
as a function of the invariant mass of the $Z$-boson pair $q$, in GeV.  Inset: 
NLO $K$-factor for $gg \to H$ as a function of the Higgs boson mass $q$, in GeV.
Bands  correspond to variations of  the renormalization and factorization scales in the interval
$ q/4 \le \mu \le q$.  The dashed line shows the $K$-factors computed for the renormalization 
and factorization scales set to $\mu = q/2$.
We  used the program MCFM \cite{mcfm} to compute  the $K$-factor for the Higgs boson 
production.
}
  \label{fig1}
\end{figure}

\section{Numerical results} 
\label{section3}

We have implemented the above formulas in a numerical {\sf Fortran} program that allows us to compute  QCD corrections 
to the top quark loop  contribution to the gluon fusion process
$pp \to ZZ$ as a function of the invariant mass of the $Z$-bosons, $q^2$.  We employ  NNPDF3.0 parton 
distribution functions \cite{Ball:2014uwa}
and use  leading order  parton distributions to compute the production cross-section at  
leading (one-loop) approximation and next-to-leading order parton distirbutions to calculate it in the two-loop approximation. 
We set the renormalization and factorization scales equal to each other. 

To assess the magnitude of QCD corrections, in the main plot of Fig.~\ref{fig1} we show the $K$-factor defined as the
 ratio of NLO and LO 
cross-sections, depending on the invariant mass of the $Z$-boson pair.  We find that the $K$-factor 
is a slowly rising function of $q^2$ and that $K \sim 1.5-1.8$ for $\mu = q/2$ for the invariant masses 
considered.   The NLO QCD corrections to the $gg \to ZZ$ process are therefore similar to what has been 
observed for other processes where gluons annihilate into colorless final states. As an illustration, 
we compare  the above results with  the NLO $K$-factors for Higgs boson production  $pp \to H$, 
shown  in the inset of Fig.~\ref{fig1}. We take the Higgs bosons
mass to be equal to the invariant mass of the $Z$-boson pair.  The $K$-factors disagree by about 
$10-15$ percent at low values of $q^2$ and agree almost perfectly at high(er) values of $q^2$.
This is in accord with the suggestion of Ref.~\cite{Bonvini:2013jha} where it was proposed 
to employ the signal $K$-factor for the description of the complete process $gg \to ZZ$ including the continuum 
contribution. 

\begin{figure}[t]
  \centering
  \includegraphics[angle=0,width=0.4\textwidth]{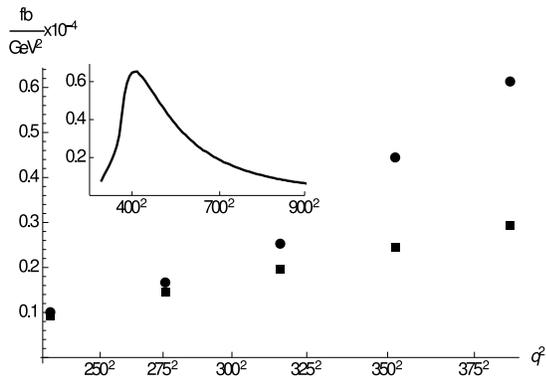}\\
  \caption{LO $pp \to ZZ$ production cross-section ( gluon fusion through top loop only) 
${\rm d}\sigma/{\rm d} q^2$ in ${\rm fb/GeV}^2$, 
as a function of the invariant mass squared of the $Z$-boson pair, $q^2$,
 in GeV${}^2$ with a top mass of 173 GeV.
We compare our cross-section, which is valid in the $m_t \to \infty$ limit, with the one implemented in MCFM, which has exact $m_t$ dependence.
The dots and squares correspond to the results from MCFM and this paper respectively.
We set the renormalization scale and the factorization scale to  $200~\mathrm{GeV}$.
The difference between the values ranges from $\sim 20\%$ to $\sim 220\%$ for the values of $q^2$ considered.
The inset shows the MCFM cross-section computed for a larger range of the invariant masses. 
} 
  \label{fig3}
\end{figure}

Finally, it is interesting to assess how the cross-section expanded in 
$1/m_t$ compares with 
the exact result. To this end, we show 
in Fig.~\ref{fig3} the leading order contribution to 
$gg \to ZZ$ with exact dependence on $m_t$ and  the $m_t \to \infty$ limit. The exact 
result is obtained using the program MCFM~\cite{mcfm}. 
We see that near threshold values of $q^2$, the two cross-sections are similar, within $\sim 20\%$.
At larger values of $q^2$ the predictions start to diverge as $1/m_t$ supressed terms become more important.
As a check of our LO cross-section we performed  a similar comparison with MCFM using a top mass 
of $m_t = 400~\mathrm{GeV}$ to simulate the $m_t \to \infty$ limit.
In this case, our calculation is within $\sim 5\%$ of the MCFM predictions for  values 
of $q^2$  between $(200~\mathrm{GeV})^2$ and $(400~\mathrm{GeV})^2$.  The inset in 
Fig.~\ref{fig3} shows the top quark loop contribution to the  
$pp \to ZZ$ cross-section with full mass dependence, as  obtained with MCFM. 
The cross-section peaks slightly above $400~{\rm GeV}$ and then starts to decrease. Our $K$-factor 
calculation is valid to the left of the peak, where the cross-section exhibits rapid growth. 
However, it can be extended beyond that by reweighting the {\it exact} $gg \to ZZ$ leading order 
partonic cross-section with $K$-factors computed in $m_t \to \infty$ limit.

\section{Conclusions}
\label{section4}

In this paper we studied QCD corrections to 
the production of a pair of $Z$-bosons in gluon fusion through loops of massive 
top quarks. This process occurs at one-loop and belongs to the class of processes 
where  two gluons annihilate into a colorless final state.  Similar to other processes 
of this type, such as  $gg \to H$ and $gg \to HH$, we find large, ${\cal O}(50-100)$ percent  
radiative corrections.   Radiative corrections of this magnitude suggest 
that the significance of gluon fusion is, perhaps, underestimated by existing  NNLO QCD computations of vector 
boson pair production in proton collisions. 

There are several avenues that are interesting to explore as the direct continuation of this work. 
First, a straightforward extension of this calculation should allow us to compute QCD radiative 
corrections to the interference 
of $gg \to ZZ$  and $gg \to H \to ZZ$ amplitudes, both on and off the mass shell of the Higgs boson.
Although such a computation will, at the moment, be restricted to top quark  contributions to 
$gg \to ZZ$, it will already give us important  information on whether or not  
the  radiative corrections to $gg \to H \to ZZ$, $gg \to ZZ$  and the interference are related. 

Second, it will be interesting to extend our calculation to include higher powers in the 
expansion of $s/m_t^2$, to estimate the impact of mass suppressed effects on QCD radiative 
corrections.  In addition, as we explained at the beginning of the paper, the effects of the 
vector coupling of $Z$-bosons  to top quarks  do not appear at leading order in the $1/m_t$ expansion which 
means that it is important to go one order higher in $1/m_t$ to fully incorporate physics of $Zt\bar t$ 
interactions into the description of the process. Of course, it is to be expected that 
since the vector coupling of $Z$-bosons to 
top quarks  is almost three times smaller than the axial coupling, the inclusion of the 
vector coupling should only lead to  small changes in the 
cross-section.

Third, it is interesting to incorporate  decays of $Z$-bosons, off-shell effects 
  and realistic selection criteria  into our calculation. This should, in principle, be straightforward
since the primary objects that we compute are the scattering amplitudes for both $gg \to ZZ^*$ and 
$gg \to ZZ^*+g$ processes. 

Finally, it is  important to combine contributions of massless quarks and 
the top quark  to $gg \to ZZ$ amplitudes in higher orders of QCD. Since the two-loop amplitudes 
for $gg \to ZZ$ with massless intermediate quarks are within reach 
\cite{Caola:2014iua,Cascioli:2014yka}, it should 
be relatively straightforward to incorporate both massless and massive quark 
loops into the description of gluon fusion contribuitions to $Z$-boson pair production.

To conclude, we described  the  calculation of NLO QCD corrections to continuum production 
of $Z$-boson pairs
in gluon fusion. The results of this  computation present first {\it direct} evidenece that the 
gluon fusion production cross-section of $Z$-bosons receives large ${\cal O}(100\%)$ QCD radiative corrections.  
Our  result  should encourage  further studies of QCD radiative corrections to weak boson pair production 
 in gluon fusion processes, mediated by massless and massive quark loops. 

{\bf Acknowledgements} 
K.M. would like to thank K.~Chetyrkin for useful conversations.
This  research is partially supported by  Karlsruhe Institute of Technology through its startup grant.

\end{document}